 \documentclass[final,3p,times,twocolumn]{elsarticle}

 \usepackage{graphicx}

\usepackage{amssymb}
\usepackage{amsthm,amsmath,amsfonts}

\usepackage{ulem}
\usepackage{color}

\begin{document}

\begin{frontmatter}

\title{$J/\psi$ yield vs. multiplicity in proton-proton collisions at the LHC}

%

\author[address1,address2]{S. Porteboeuf}
\ead{sarah.porteboeuf@clermont.in2p3.fr}

\author[address1]{R. Granier de Cassagnac}
\ead{raphael@in2p3.fr}

\address[address1]{Laboratoire Leprince-Ringuet (LLR), \'Ecole Polytechnique, IN2P3-CNRS, Palaiseau, France}
\address[address2]{Laboratoire de Physique Corpusculaire (LPC), Clermont Universit\'e, Universit\'e Blaise Pascal, CNRS-IN2P3, Clermont-Ferrand, France}

\begin{abstract}
We address the question of understanding the production of $J/\psi$ particles regarding the global underlying event in proton-proton collisions. To do so, we propose to look at a new observable: the $J/\psi$ production as a function of the charged particles multiplicity of the event. We demonstrate the interest for an experimental measurement by varying the model of multiple interactions in the PYTHIA generator.
\end{abstract}


\end{frontmatter}


\section{Motivation}
\label{motivation}

Understanding the production of the $J/\psi$ particle in proton-proton collisions is a complex issue \cite{Lansberg_review,Kramer_review}. It becomes even more complex when one considers the production of a $J/\psi$ in a complete event containing interplaying soft and hard components produced simultaneously. In this first section, we provide motivations for studying quarkonia production regarding the complete proton+proton event. We emphasize that proton-proton collisions are not trivial, especially in the regime of high multiplicity events.

\begin{figure}[h]
	\centering
	\includegraphics[width=1.0\columnwidth]{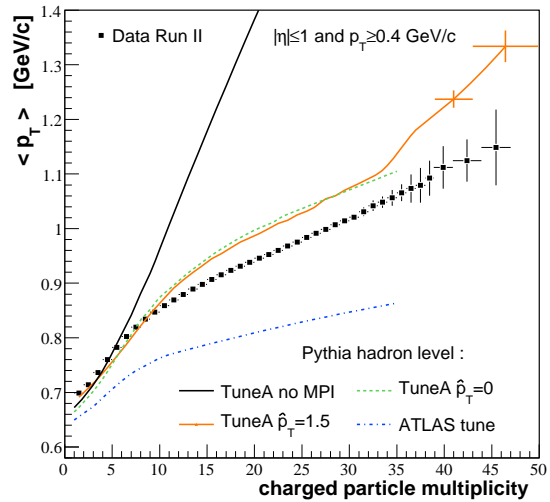}
	\caption{$\langle p_T\rangle$ as a function of multiplicity. The data points (black squares) are compared to different event generator simulations from PYTHIA. In PYTHIA: improvement of the multiple interactions scheme leads to a better description of the CDF data~\cite{CDF_mult_PYTHIA}.}
	\label{CDF}
\end{figure}

\smallskip
Very few observables have been studied as a function of the charged particles multiplicity. The CDF collaboration conducted such a study for the mean transverse momentum $\langle p_T\rangle$, as illustrated on Fig~\ref{CDF}. An important raise of $\langle p_T\rangle$ is seen as a function of multiplicity. This can result from several elementary {\it multiple interactions} occurring in parallel in growing multiplicity events. In these events, one can also expects non trivial phenomenon, such as collective effects, as mentioned later. Thus, there is no theoretical clear picture for the explanation of $\langle p_T\rangle$ versus multiplicity, but multiple interactions clearly play an important role.  Another feature that helps to understand the importance of multiple interactions, is the fact that, for proton-proton collisions, the jet cross section computed in the framework of perturbative Quantum Chromodynamics (pQCD), becomes larger than the total cross section: there is thus more than one elementary collision in high multiplicity events~\cite{Wang}.

\smallskip
On Fig.~\ref{CDF}, authors compared this observable with the event generator PYTHIA~\cite{PYTHIA_Manuel}. These event generator is based on pQCD and assume the factorization scheme. At the time the CDF data, Run I, was released, none of the models were able to describe the data~\cite{CDF_mult}. After improving its treatment of multiple interactions, PYTHIA finally describes better these data with the tune $\hat{A}$ ($\hat{p}_{\perp}=1.5$)~\cite{CDF_mult_PYTHIA}. A good understanding of high multiplicity phenomena is thus important to completely describe events with hard and soft components, and possible interactions between the two. It is to be noted that collective effects in proton-proton collisions can also contribute to explaining this observable, as for instance accounted in the EPOS event generator~\cite{perugia}.

\smallskip
 The only other multiplicity dependence study that we are aware of is from one of the first UA1 paper where the authors studied the transverse energy $E_T$ as a function of multiplicity~\cite{UA1_first}. Besides these CDF and UA1 observations, and a recent CMS result that will be mentioned hereafter, high multiplicity phenomena in proton-proton collisions are not studied. On the opposite, in heavy-ions collisions, studying observables as a function of the (very high) multiplicity is a key concept and a very common tool to study the Quark Gluon Plasma (QGP).


\section{$J/\psi$ production in complete events}

In standard proton-proton collisions, $J/\psi$ are produced through hard processes. Hard processes can be understood in the framework of pQCD, where in the limit of high momentum transfer $Q^2$, a hard component ($2\to2$ or $2\to1$ processes) can be factorized out from the soft component, as predicted by the factorization theorem \cite{pQCD, factorization}.
$J/\psi$ are very particular among the hard probes, because they first imply the production of a massive charm quark-antiquark pair, followed by its binding into a charmonia ($c\bar{c}$) state. Nowadays, the exact charmonia production scheme is not clear and is debated within the community (for reviews on this aspect, see~\cite{Lansberg_review, Kramer_review}). Furthermore, no models describe the underlying event associated to a $J/\psi$ production, and one needs to be aware of this lack of description, when dealing with event generators. This problem was for instance raised in February 2010 during a workshop on quarkonia production at CERN~\cite{workshop_quarkonium}. New observables are then needed to start understanding and constraining the interplay of soft and hard aspects of proton-proton collisions, the importance of multiple interactions in the production of hard processes and the possibility of new phenomena such as collective effects.


The simple new observable we propose to look at is the $J/\psi$ yield versus the underlying event multiplicity. The basic idea is to compare the multiplicity of events containing a $J/\psi$ to the multiplicity of Minimum Bias events.
The physics that we want to address here is the interplay between the hard ($J/\psi$) and the soft (underlying) components in a complete event, the influence of one over the other, in the case of quarkonia production such as the $J/\psi$. The question is: do the underlying event and the multiple interactions scheme matter in the production of hard processes, such as a $J/\psi$? Though we limit ourselves to the case of $J/\psi$, we think the complete understanding of global events will only arise by studying as many quantities as possible, including various hard probes.

\section{First simple PYTHIA study}

\smallskip
To test the relevance of our new observable, we conducted a simple study at the event generator level. To do so, we needed an event generator that is able to produce simultaneously the $J/\psi$ and the corresponding underlying event. PYTHIA is one of the most developed event generator having the two functionalities. Other event generators have more sophisticated description of the soft component (HERWIG \cite{HERWIG}, EPOS \cite{EPOS}), but none of them include the production of quarkonia.
Another possibility could be the CASCADE event generator \cite{cascade}, this could be explored.

\smallskip
We generate two uncorrelated sets of PYTHIA events. In the first set, it is required to produce a $J/\psi$ in each event (ISUB=86), through the color singlet model used by default in PYTHIA~\cite{PYTHIA_Manuel}. The second set is made of Minimum Bias ($MB$) events (MSEL=2). The multiplicity distribution is determined for the two sets and their ratio ($J/\psi \div MB$) is computed, then normalized for the $J/\psi$ and $MB$ cross sections, estimated by PYTHIA.

\smallskip
It is to be noted here, that our $J/\psi$ are directly produced and do not come from $B$, $\chi_c$ or $\psi'$ decays.

\begin{figure}[!htp]
	\centering
	\includegraphics[width=1.1\columnwidth]{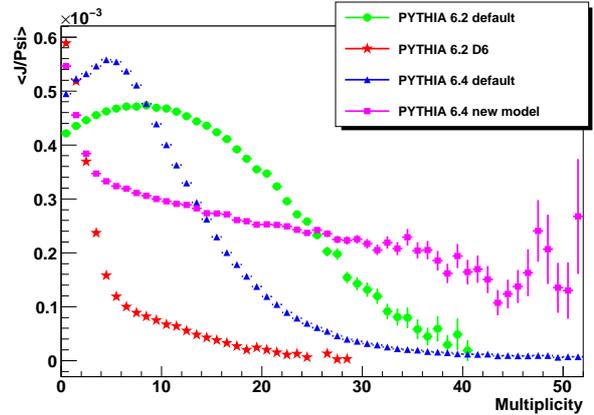}
	\caption{PYTHIA Study of $J/\psi$ production as a function of multiplicity, in proton-proton collisions at 10 Tev for different PYTHIA settings. For the multiplicity, only charged particles with a $p_{T}>0.9$ GeV/c and $|y|<2.4$ are considered. }
	\label{PYTHIA}
\end{figure}

\smallskip
Fig.~\ref{PYTHIA} shows this first result with a million generated events for each set, for proton-proton collisions at 10~TeV. Here, $Multiplicity$ stands for the number of charged particles with $p_{T}>0.9$~GeV/$c$ and pseudorapidity lower than 2.4, corresponding to the acceptance of the CMS tracker. In all the PYTHIA tunes we considered, the $J/\psi$ production mechanism is the default one, meaning the Color Singlet model, but one should also looked at different implementations of the production mechanism for other birth processes (such as the Color Octet mechanism).

\smallskip
The four curves refer to different PYTHIA settings, mostly implying variation of the multiple interactions model.

\begin{itemize}
\item The green closed circle symbols account for basic PYTHIA, version 6.2, with all parameters set as default (MSTP(82)=1).
In this simple model all interactions are supposed to be independent. The number of parton-parton interactions in a proton-proton collision follows a Poissonian law, with the mean number of interactions $\bar{n}=\sigma_{\textrm{hard}}/\sigma_{\textrm{inelastic}}$.
 Here, the hard process is the first one, the hardest one, the successive multiple interactions being then ordered by their hardness. There is no initial state radiations for interactions in the multiple scattering and no color connection between different ladders.

\item For the red closed star symbols, the PYTHIA 6.2 tune D6 is used. This is the best tune for Tevatron data. With the PYTHIA parameter MSTP(82)=4, the model for multiple interactions is improved. Hadrons are considered to be extended objects. The collision has an {\it impact parameter} which modifies the number of multiple interactions. When the protons overlap a lot, there is more chance to have multiple interactions than when there is less.

\item The blue closed triangle symbols is PYTHIA 6.4 all parameters by default, the model  for multiple interactions is based on the model in PYTHIA 6.2, with a modification of the default values for different parameters (hard scale). This is an intermediate version between PYTHIA 6.2 D6 and the new model.

\item For the purple closed square symbols, a new multiple interactions model was developed to be able to describe CDF data on $\langle p_T\rangle$ (Fig.~\ref{CDF}). In this new model, based on the previous one for the determination of the number of interactions, initial state radiations are implemented with color re-connections between the strings produced by different interactions. The first interaction is still the hardest one. With this new model, more particles are produced in event with a lot of multiple interactions, color re-connections are an important feature for the structure of the whole event.
\end{itemize}

A detailed description of the physics implied in the different PYTHIA models for multiple interactions can be found in~\cite{PYTHIA_Manuel}.

\smallskip
The naive expectation is that, if the production of a $J/\psi$ has nothing to do with multiple interactions, we should observe a flat ratio on Fig.~\ref{PYTHIA}. It is clearly not the case, in all PYTHIA configurations. Furthermore, the results depends strongly on the PYTHIA model of multiple interactions assumed. Quite unexpected is the shape of the strongly decreasing curves we observe, especially in the most modern tunes. It is to be noted that the multiple interactions scheme in PYTHIA is still under study and the production of quarkonia is not well constrained, especially with the final state radiation, and the production mechanism.

\smallskip
In any case, the physical meaning (in PYTHIA) of the different curves is hard to understand. A detailed study of all PYTHIA parameters would be necessary to identify what mechanism is responsible in PYTHIA for each kind of multiple interactions models, (initial sate interaction, final state interaction, pQCD scales, color re-connection...).

\smallskip
More important is to study this observable in real data, for instance in the 7~TeV proton-proton data collected at LHC, and to compare the experimental behavior to the simulated ones. We do not expect PYTHIA to quantitatively predict such a new observable, for which the event generator was not built, but rather the experimental measurement to help tuning PYTHIA.

\smallskip
To interpret this future measurement, an important problem will be its normalization. For corresponding study in heavy-ions collisions, one compares to proton-proton collisions, scaled by the number of equivalent proton-proton collisions. Here we are looking only at proton-proton collisions: there is no such underlying elementary collisions concept. The normalization is then not clear and can be debated.

\smallskip
A possibility would be to compare with open charm. Open charm are produced in standard $2\to2$ hard processes. If $J/\psi$ has the same behavior as a standard hard processes from the multiplicity point of view, we should see a similar behavior.
One can also think to look at different $p_T$ classes of the $J/\psi$ to look for eventual modification of the trend looking at hardest particles.
Finally, in a more general study, we would recommend to look at any observable, one could think of, as a function of multiplicity. Taking hadrons or jets for example, one then could use another event generator than PYTHIA, with a different feature of multiple interactions, and collective phenomenon, and here, one could think of EPOS~\cite{EPOS}, that does not include, massive quarks and quarkonia production.

\section{Conclusion}
This first generator level study of $J/\psi$ production as a function of multiplicity shows that the production of $J/\psi$ in a complete event with the corresponding underlying event and multiple interactions scenario does not seem to be under control in the PYTHIA event generator. Our recommendation is to perform this study with the new LHC data at 7~TeV in proton-proton collisions. The ALICE, ATLAS, CMS and LHCb experiments have complementary acceptance. For instance, CMS can measure $J/\psi$ with $|y|<2.4$ but missing the lower $p_T$ part of the spectrum at midrapidity, while ALICE will measured $J/\psi$ in rapidity ranges of $|y|<0.8$ and $2.5<y<4$, at all $p_T$. This measurement should be performed in the four experiments and compared. More generally, a detailed study for all observables should be performed regarding the multiplicity.

Recently, the CMS experiment showed an interesting and unexpected feature of high multiplicity events in proton-proton collisions \cite{CMS_ridge}. The observed long range correlations are hard to interpret, but already tells us that high multiplicity proton-proton collisions are not trivial.

\section*{Acknowledgments}

Sarah Porteboeuf wants to thanks the ReteQuarkonii Network and the I3HP2 European Program which provided the grant supporting this work.





\bibliographystyle{elsarticle-num}
\bibliography{biblio_porteboeuf_quarkonium2010}

\begin{thebibliography}{10}
\expandafter\ifx\csname url\endcsname\relax
  \def\url#1{\texttt{#1}}\fi
\expandafter\ifx\csname urlprefix\endcsname\relax\def\urlprefix{URL }\fi
\expandafter\ifx\csname href\endcsname\relax
  \def\href#1#2{#2} \def\path#1{#1}\fi

\bibitem{Lansberg_review}
J.~P. Lansberg, {$J/\psi$, $\psi$ ' and $\upsilon$ production at hadron
  colliders: A Review}, Int. J. Mod. Phys. A21 (2006) 3857--3916.
\newblock \href {http://arxiv.org/abs/hep-ph/0602091}
  {\path{arXiv:hep-ph/0602091}}, \href
  {http://dx.doi.org/10.1142/S0217751X06033180}
  {\path{doi:10.1142/S0217751X06033180}}.

\bibitem{Kramer_review}
M.~Kramer, 1, {Quarkonium production at high-energy colliders}, Prog. Part.
  Nucl. Phys. 47 (2001) 141--201.
\newblock \href {http://arxiv.org/abs/hep-ph/0106120}
  {\path{arXiv:hep-ph/0106120}}, \href
  {http://dx.doi.org/10.1016/S0146-6410(01)00154-5}
  {\path{doi:10.1016/S0146-6410(01)00154-5}}.

\bibitem{Wang}
X.-N. Wang, M.~Gyulassy, {A Systematic study of particle production in $p$ +
  $p$ (anti-p) collisions via the HIJING model}, Phys. Rev. D45 (1992)
  844--856.
\newblock \href {http://dx.doi.org/10.1103/PhysRevD.45.844}
  {\path{doi:10.1103/PhysRevD.45.844}}.

\bibitem{CDF_mult_PYTHIA}
T.~Aaltonen, et~al., {Measurement of Particle Production and Inclusive
  Differential Cross Sections in $p\bar{p}$ Collisions at $\sqrt{s}=1.96$ TeV},
  Phys. Rev. D79 (2009) 112005.
\newblock \href {http://arxiv.org/abs/0904.1098} {\path{arXiv:0904.1098}},
  \href {http://dx.doi.org/10.1103/PhysRevD.79.112005}
  {\path{doi:10.1103/PhysRevD.79.112005}}.

\bibitem{PYTHIA_Manuel}
T.~Sjostrand, S.~Mrenna, P.~Z. Skands, {PYTHIA 6.4 Physics and Manual}, JHEP 05
  (2006) 026.
\newblock \href {http://arxiv.org/abs/hep-ph/0603175}
  {\path{arXiv:hep-ph/0603175}}, \href
  {http://dx.doi.org/10.1088/1126-6708/2006/05/026}
  {\path{doi:10.1088/1126-6708/2006/05/026}}.

\bibitem{CDF_mult}
D.~E. Acosta, et~al., {Soft and hard interactions in $p\bar{p}$ collisions at
  $\sqrt{s}=$ 1800-GeV and 630-GeV}, Phys. Rev. D65 (2002) 072005.
\newblock \href {http://dx.doi.org/10.1103/PhysRevD.65.072005}
  {\path{doi:10.1103/PhysRevD.65.072005}}.

\bibitem{perugia}
P.~Bartalini, (ed.~), et~al., {Proceedings of the First International Workshop
  on Multiple Partonic Interactions at the LHC (MPI08)}\href
  {http://arxiv.org/abs/1003.4220} {\path{arXiv:1003.4220}}.

\bibitem{UA1_first}
G.~Arnison, et~al., {Some Observations on the First Events Seen at the CERN
  Proton - anti-Proton Collider}, Phys. Lett. B107 (1981) 320--324.
\newblock \href {http://dx.doi.org/10.1016/0370-2693(81)90839-X}
  {\path{doi:10.1016/0370-2693(81)90839-X}}.

\bibitem{pQCD}
J.~C. Collins, D.~E. Soper, G.~F. Sterman, {Factorization of Hard Processes in
  QCD}, Adv. Ser. Direct. High Energy Phys. 5 (1988) 1--91.
\newblock \href {http://arxiv.org/abs/hep-ph/0409313}
  {\path{arXiv:hep-ph/0409313}}.

\bibitem{factorization}
J.~C. Collins, D.~E. Soper, G.~F. Sterman, {Factorization for Short Distance
  Hadron - Hadron Scattering}, Nucl. Phys. B261 (1985) 104.
\newblock \href {http://dx.doi.org/10.1016/0550-3213(85)90565-6}
  {\path{doi:10.1016/0550-3213(85)90565-6}}.

\bibitem{workshop_quarkonium}
{Quarkonium production at the LHC, 19th February 2010, CERN,
  http://indico.cern.ch/conferenceDisplay.py?confId=78640}.

\bibitem{HERWIG}
G.~Corcella, et~al., {HERWIG 6.5: an event generator for Hadron Emission
  Reactions With Interfering Gluons (including supersymmetric processes)}, JHEP
  01 (2001) 010.
\newblock \href {http://arxiv.org/abs/hep-ph/0011363}
  {\path{arXiv:hep-ph/0011363}}.

\bibitem{EPOS}
K.~Werner, {The hadronic interaction model EPOS}, Nucl. Phys. Proc. Suppl.
  175-176 (2008) 81--87.
\newblock \href {http://dx.doi.org/10.1016/j.nuclphysbps.2007.10.012}
  {\path{doi:10.1016/j.nuclphysbps.2007.10.012}}.

\bibitem{cascade}
H.~Jung, G.~P. Salam, {Hadronic final state predictions from CCFM: The hadron-
  level Monte Carlo generator CASCADE}, Eur. Phys. J. C19 (2001) 351--360.
\newblock \href {http://arxiv.org/abs/hep-ph/0012143}
  {\path{arXiv:hep-ph/0012143}}, \href
  {http://dx.doi.org/10.1007/s100520100604} {\path{doi:10.1007/s100520100604}}.

\bibitem{CMS_ridge}
{Observation of Long-Range Near-Side Angular Correlations in Proton-Proton
  Collisions at the LHC}, JHEP 09 (2010) 091.
\newblock \href {http://arxiv.org/abs/1009.4122} {\path{arXiv:1009.4122}},
  \href {http://dx.doi.org/10.1007/JHEP09(2010)091}
  {\path{doi:10.1007/JHEP09(2010)091}}.

\end{thebibliography}






\end{document}